\documentclass[10pt,letterpaper]{article}
\usepackage{opex3}
\usepackage{color}
\usepackage{amsfonts}
\usepackage{amsmath}
\usepackage{amssymb}
\usepackage{graphicx}
\usepackage{cite}

\providecommand{\U}[1]{\protect\rule{.1in}{.1in}}

\begin{document}
\title{Diffraction manipulation by four-wave mixing}

\author{Itay Katzir$^{1,*}$, Amiram Ron$^1$, and Ofer Firstenberg$^{2}$}

\address{$^1$Department of Physics, Technion-Isreal Institute of Technology, Haifa 32000, Israel\\
$^2$Department of Physics of Complex Systems, Weizmann Institute of Science, Rehovot 76100, Israel}
\email{$^*$itaykatzir@gmail.com} 

\begin{abstract}
We suggest a scheme to manipulate paraxial diffraction by utilizing the
dependency of a four-wave mixing process on the relative angle between the
light fields. A microscopic model for four-wave mixing in a $\Lambda$-type
level structure is introduced and compared to recent experimental data. We
show that images with feature size as low as 10 $\mu$m can propagate with very
little or even negative diffraction. The mechanism is completely different from that conserving the shape of spatial solitons in nonlinear media, as here diffraction is suppressed for arbitrary spatial profiles. At the same time, the gain inherent to the nonlinear process prevents loss and
allows for operating at high optical depths. Our scheme does not rely on
atomic motion and is thus applicable to both gaseous and solid media.
\end{abstract}



\section{Introduction}
The diffraction of light during propagation in free space is a fundamental and
generally unavoidable physical phenomenon. Diffracting light beams
do not maintain their intensity distribution in the plane transverse to the
propagation direction, unless belonging to a particular class of
non-diffracting (Bessel) beams \cite{DurninJOSA1987}. In nonuniform media,
waveguiding is possible for specific spatial modes
\cite{AgarwalPRA2000,TruscottPRL1999}, or equivalently arbitrary images may
revive after a certain self-imaging distance \cite{ChengOL2007}. However in
such waveguides, the suppression of diffraction for multimode profiles is not
trivial, as each transverse mode propagates with a different propagation
constant or group velocity, resulting in spatial dispersion of the profile.

Recently, a mechanism was suggested and demonstrated for manipulating the diffraction of arbitrary images imprinted on a light
beam for arbitrary propagation distances \cite{FirstenbergPRL2009,FirstenbergNP2009}. The technique is based on
electromagnetically induced transparency (EIT) \cite{FleischhauerRMP2005} in a
thermal atomic gas. Unlike other methods utilizing EIT
\cite{AgarwalPRA2000,TruscottPRL1999,ChengOL2007,MoseleyPRL1995,MitsunagaPRA2000,WilsonGordonPRA1998,LukinPRL2005_strong_confinement,chengPRA2005,VengalattorePRL2005,TarhanOL2007,HongPRL2003,FriedlerOL2005,GomezReinoAO1982}%
, which rely on spatial non-uniformity, this technique prescribes
non-uniformity in $k_{\perp}$ space. Here, $k_{\perp}$ denotes the transverse
wave vectors, \textit{i.e.}, the Fourier components of the envelope of the
field in the transverse plane, which is the natural basis for paraxial
diffraction. The technique of Refs.~\cite{FirstenbergPRL2009,FirstenbergNP2009} relies on the diffusion of the
atoms in the medium and on the resulting diffraction-like optical response. However, the resolution limit of such motional-induced diffraction in currently available experimental conditions is on the order of 100 $\mu$m, preventing it from being of much practical use. Higher resolution requires
a denser atomic gas, in which strong absorption is unavoidable due to
imperfect EIT conditions. Very recently, Zhang and Evers proposed to
circumvent the absorption by generalizing the model of motional-induced
diffraction to a four-wave mixing (FWM) process in combination with EIT
\cite{EversPRA2014}. The FWM process further allows the frequency conversion
of the image and increases the available resolution.

In this paper, we propose a scheme to manipulate diffraction using FWM
\cite{GarmirePRL1966,KelleyPRL1966,BoydPRA1981} \emph{without} the need for
motional-induced diffraction. The mechanism we study originates from phase
matching in $k_{\perp}$ space and does not require a gaseous medium; it is
therefore directly applicable to solid nonlinear media. For our model to be
general and to accommodate motional-broadening mechanisms (not important in
solids), we here still concentrate on describing atomic gases and validate our
model against relevant experiments. The inherent gain of the FWM process
allows us to improve the spatial resolution by working with relatively higher
gas densities while avoiding loss due to absorption.

In Sec.~\ref{s_theory}, we introduce a microscopic model of FWM in a $\Lambda$ system,
based on Liouville-Maxwell equations, similar to the one used in Ref.~\cite{HaradaPRA013809}. In Sec.~\ref{s_experiments}, we compare the model to recent experimental results of
FWM in hot vapor \cite{HaradaPRA013809,LettArxiv2007}. We use our model in
Sec.~\ref{s_diffraction} to show that, with specific choice of frequencies, the
$k_{\perp}$ dependency of the FWM process can be used to eliminate the
diffraction of a propagating light beam. We also present a demonstration of
negative diffraction, implementing a paraxial version of a negative-index lens
\cite{Veselago1968}, similar to the one in Ref.~\cite{FirstenbergNP2009} but with positive gain and higher resolution.
Finally, we analyze the resolution limitation of our scheme and propose ways
to enhance it. We show that, for cold atoms at high densities ($\sim$10$^{12}$
cm$^{-3}$), diffraction-less propagation of an image with a resolution of
$\sim$10 $\mu$m can be achieved.%

\begin{figure}[ptb]
\centering\includegraphics[width=12cm]
{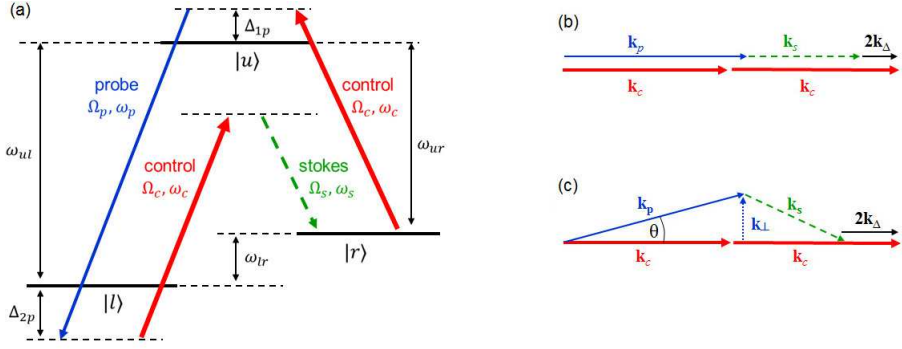}%
\caption{(a) Four-wave mixing in a three-level $\Lambda$ system ($\left\vert
u\right\rangle $, $\left\vert l\right\rangle $, \ and $\left\vert
r\right\rangle $). $\Omega_{i}$\ with $i=c,p,s$\ are the Rabi frequencies of
the fields. The phase-matching conditions are shown for (b) collinear and (c)
non-collinear propagation. The phase mismatch scalar is $2k_{\Delta}=\left(
2\boldsymbol{k}_{c}-\boldsymbol{k}_{p}-\boldsymbol{k}_{s}\right)
\mathbf{\hat{z}}$.}\label{FIG1}
\end{figure}

\section{Theory\label{s_theory}}

\subsection{Model}

We consider an ensemble of three-level atoms in a $\Lambda$ configuration
depicted in Fig.~\ref{FIG1}a. The atomic states are denoted as $\left\vert u\right\rangle $,
$\left\vert l\right\rangle $, and $\left\vert r\right\rangle ,$ for the up,
left, and right states, and the optical transition
frequencies $\omega_{ul}$ and $\omega_{ur}$ are assumed to be much larger than the ground-state splitting $\omega_{lr}=\omega_{ul}-\omega_{ur}$. The atom interacts with a weak 'probe' and two 'control' electromagnetic fields, propagating in time $t$ and space $\mathbf{r}$,%
\begin{equation}%
\begin{array}
[c]{c}%
\mathbf{E}_{cl}\left(  \mathbf{r},t\right)  =(\hslash/\mu
)\boldsymbol{\varepsilon}_{cl}\Omega_{c}\left(  \mathbf{r},t\right)
e^{-i\omega_{c}t}e^{ik_{0}^{c}z},\\
\mathbf{E}_{cr}\left(  \mathbf{r},t\right)  =(\hslash/\mu
)\boldsymbol{\varepsilon}_{cr}\Omega_{c}\left(  \mathbf{r},t\right)
e^{-i\omega_{c}t}e^{ik_{0}^{c}z},\\
\mathbf{E}_{p}\left(  \mathbf{r},t\right)  =(\hslash/\mu
)\boldsymbol{\varepsilon}_{p}\Omega_{p}\left(  \mathbf{r},t\right)
e^{-i\omega_{p}t}e^{ik_{0}^{p}z}.
\end{array}\label{eq:Eq1}
\end{equation}
To simplify the formalism, the same dipole moment is assumed for the two optical
transitions $\mu=\mu_{ul}=\mu_{ur}$, and the two control fields differ only in their polarizations. Here $\omega_{i}$ are the frequencies of the 'probe' ($i=p$) and the 'control' fields ($i=c$); $k_{0}^{i}\equiv\omega_{i}/c$ are the wave vectors
in the case of plane waves, otherwise they are carrier wave-vectors; and
$\Omega_{i}\left(  \mathbf{r},t\right)$ are the slowly varying envelopes of
the the Rabi frequencies, satisfying $\left\vert \partial_{t}^{2}\Omega
_{i}\left(  \mathbf{r},t\right)  \right\vert \ll\left\vert \omega_{i}%
\partial_{t}\Omega_{i}\left(  \mathbf{r},t\right)  \right\vert $ and
$\left\vert \partial_{z}^{2}\Omega_{i}\left(  \mathbf{r},t\right)  \right\vert
\ll\left\vert k_{i}^{0}\partial_{z}\Omega_{i}\left(  \mathbf{r},t\right)
\right\vert$.  The polarization
vectors of the fields are $\boldsymbol{\varepsilon}_{p}$, $\boldsymbol{\varepsilon}_{cl}$, and $\boldsymbol{\varepsilon}_{cr}$. The strong control and weak probe fields stimulate a weak
classical 'Stokes' field (or 'conjugate') at a frequency $\omega_{s}%
=2\omega_{c}-\omega_{p},$%
\begin{equation}
\mathbf{E}_{s}\left(  \mathbf{r},t\right)  =(\hslash/\mu
)\boldsymbol{\varepsilon}_{s}\Omega_{s}\left(  \mathbf{r},t\right)
e^{-i\omega_{s}t}e^{ik_{0}^{s}z}.
\end{equation}

To further simplify the analysis, we assume a single relaxation rate $\Gamma$ between the excited and ground levels and define the complex rates $\gamma_{cr}=\Gamma-i\left(  \omega_{c}-\omega_{ur}\right)$ and $\gamma_{cl}=\Gamma-i\left(  \omega_{c}-\omega_{ul}\right)$ for each of the optical transitions. Within the ground state, we consider a population relaxation with symmetric rates $\Gamma_{l\leftrightarrow r}$ and a decoherence with a rate $\Gamma_{lr}$. In a frame rotating with the control frequency $\omega_{c}$, the equations of
motion for the local density-matrix $\rho\left(  \mathbf{r},t\right)  $ are
better written in terms of the slowly-varying density-matrix $R\left(
\mathbf{r},t\right)  ,$ where $R_{u,j}\left(  \mathbf{r},t\right)  =\rho
_{u,j}\left(  \mathbf{r},t\right)  e^{i\omega_{c}t-ik_{0}^{c}z}$ for $j=l,r$
and $R_{\alpha,\alpha^{\prime}}\left(  \mathbf{r},t\right)  =\rho
_{\alpha,\alpha^{\prime}}\left(  \mathbf{r},t\right)  $ for all other matrix
elements,%
\begin{align}
\frac{\partial}{\partial t}{\small R}_{l,l}  & {\small =}{\small -2}%
\operatorname{Im}(\hat{P}^{\ast}R_{u,l}){\small +\Gamma}_{l\leftrightarrow
r}{\small (R}_{r,r}{\small -R}_{l,l}{\small )+\Gamma R}_{u,u}\nonumber\\
\frac{\partial}{\partial t}R_{r,r}  & =-2\operatorname{Im}(\hat{S}^{\ast
}R_{u,r})-\Gamma_{l\leftrightarrow r}(R_{r,r}-R_{l,l})+\Gamma R_{u,u}%
\nonumber\\
\frac{\partial}{\partial t}R_{u,u}  & =2\operatorname{Im}(\hat{P}^{\ast
}R_{u,l})+2\operatorname{Im}(\hat{S}^{\ast}R_{u,r})-2\Gamma R_{u,u}\nonumber\\
\frac{\partial}{\partial t}R_{r,l}  & =i\hat{S}^{\ast}R_{u,l}-i\hat{P}%
R_{u,r}^{\ast}-(\Gamma_{lr}+i\omega_{lr})R_{r,l}\nonumber\\
\frac{\partial}{\partial t}R_{u,l}  & =-i\hat{P}\left(  R_{u,u}-R_{l,l}%
\right)  +i\hat{S}R_{r,l}+\gamma_{cl}^{\ast}R_{u,l}\nonumber\\
\frac{\partial}{\partial t}R_{u,r}  & =-i\hat{S}\left(  R_{u,u}-R_{r,r}%
\right)  +i\hat{P}R_{r,l}^{\ast}+\gamma_{cr}^{\ast}R_{u,r}.\label{e5}%
\end{align}
Here
\begin{equation}
\hat{P}  \equiv\Omega_{p}\left(  \mathbf{r},t\right)  e^{-i\left(
\delta_{\omega}t-\delta_{k}z\right)  }+\Omega_{c} \quad \textrm{and} \quad
\hat{S}  \equiv\Omega_{s}\left(  \mathbf{r},t\right)  e^{i\left(
\delta_{\omega}t-\delta_{k}z\right)  }+\Omega_{c}
\end{equation}
are interference fields, and $\delta_{\omega}%
=\omega_{p}-\omega_{c}=\omega_{c}-\omega_{s}$ and $\delta_{k}=k_{0}^{p}%
-k_{0}^{c}$ $=k_{0}^{c}-k_{0}^{s}$ are detuning parameters.

Finally assuming non-depleted control fields, constant in time and space $\Omega
_{c}\left(  \mathbf{r},t\right)  =\Omega_{c},$ we complete the description of
the atom-field interaction with the propagation equations under the envelope
approximation for the probe field
\begin{subequations}
\label{e6}%
\begin{equation}
\left(  \frac{\partial}{\partial z}+\frac{1}{c}\frac{\partial}{\partial
t}+\frac{i\nabla_{\perp}^{2}}{2q}\right)  \Omega_{p}\left(  \mathbf{r}%
,t\right)  =igR_{u,l}\left(  \mathbf{r},t\right)  e^{i(\delta_{\omega}%
t-\delta_{k}z)}%
\end{equation}
and the Stokes field%
\begin{equation}
\label{e6}\left(  \frac{\partial}{\partial z}+\frac{1}{c}\frac{\partial
}{\partial t}-\frac{i\nabla_{\perp}^{2}}{2q}\right)  \Omega_{s}\left(
\mathbf{r},t\right)  =igR_{u,r}\left(  \mathbf{r},t\right)  e^{-(\delta
_{\omega}t-\delta_{k}z)},
\end{equation}
where $\nabla_{\perp}^{2}\equiv$ $\partial^{2}/\partial x^{2}+\partial
^{2}/\partial y^{2}$ the transverse Laplacian, $g=2\pi N\left\vert
\mu\right\vert ^{2}q/\hslash$ the coupling strength proportional to the atomic
density $N$, and $q\equiv\left\vert k_{0}^{c}\right\vert \approx\left\vert
k_{0}^{p}\right\vert \approx\left\vert k_{0}^{s}\right\vert $. To obtain Eqs.~(\ref{e6}), we neglected the second-order $t$ and $z$ derivatives of the envelopes.

\subsection{Steady-state solution}

The evolution of the fields is described by a set of non-linear, coupled
differential equations for the density matrix elements $R_{\alpha
,\alpha^{\prime}}$ and the weak fields $\Omega_{p}$ and $\Omega_{s}$ [Eqs.~(\ref{e5})-(\ref{e6})], which require further approximations in order to be solved
analytically. Most importantly, we assume the proximity to two-photon resonance, such that $\delta_{\omega}$ is on the order of the ground-state frequency splitting
$\omega_{lr}$ and much larger than any detuning, Rabi frequency, or pumping
rate in the system. Other assumptions are detailed in the appendix, where we derive the steady state of the system to first order in the weak fields,
\end{subequations}
\begin{equation}
R_{\alpha,\alpha^{\prime}}\simeq R_{\alpha,\alpha^{\prime}}^{(0)}%
+R_{\alpha,\alpha^{\prime}}^{(1)},\label{RR0R1}%
\end{equation}
with $R_{\alpha,\alpha^{\prime}}^{(0)}$ and $R_{\alpha,\alpha^{\prime}}^{(1)}$
being the zero- and first- order steady-state solutions. Plugging Eqs.~(\ref{RPN})-(\ref{Pur1}) and (\ref{RR0R1}) for $R_{u,r}$ and $R_{u,l}$ into
the propagation equations (\ref{e6}) and discarding terms rotating at
$\delta_{\omega}$ and $2\delta_{\omega}$, we obtain the well-known FWM form
including paraxial diffraction,
\begin{subequations}
\label{4wm-1}%
\begin{align}
\left(  \frac{\partial}{\partial z}-i\frac{1}{2q}\nabla_{\perp}^{2}\right)
\Omega_{p}\left(  \mathbf{r}\right)   & =A\Omega_{p}\left(  \mathbf{r}\right)
+B\Omega_{s}^{\ast}\left(  \mathbf{r}\right), \\
\left(  \frac{\partial}{\partial z}+i\frac{1}{2q}\nabla_{\perp}^{2}\right)
\Omega_{s}^{\ast}\left(  \mathbf{r}\right)   & =C\Omega_{p}\left(
\mathbf{r}\right)  +D\Omega_{s}^{\ast}\left(  \mathbf{r}\right)  ,
\end{align}
\end{subequations}
where
\begin{align}
A  & \equiv-\alpha_{p}+\frac{\beta_{p}}{\gamma_{pl}\gamma_{0}}\left\vert
\Omega_{c}\right\vert ^{2},\ \ \ B\equiv\frac{\beta_{s}}{\gamma_{pl}\gamma
_{0}}\left\vert \Omega_{c}\right\vert ^{2},\label{a3}\\
C  & \equiv\frac{\beta_{p}}{\gamma_{sr}^{\ast}\gamma_{0}}\left\vert \Omega
_{c}\right\vert ^{2},\ \ \ \ \ \ \ \ \ \ \ \ D\equiv-\alpha_{s}^{\ast}%
+\frac{\beta_{s}}{\gamma_{sr}^{\ast}\gamma_{0}}\left\vert \Omega
_{c}\right\vert ^{2}.\nonumber
\end{align}
Here $\alpha_{j}=g\left(  n_{l}/\gamma_{jl}+n_{r}/\gamma_{jr}\right)  $ are
the linear absorption coefficients of the probe ($j=p$) or Stokes ($j=s$)
fields, with $n_{i}\equiv R_{i,i}^{(0)}$\ the populations of the$\ i=r,l$
levels. $\beta_{p}=g\left(  n_{l}/\gamma_{pl}+n_{r}/\gamma_{cr}^{\ast}\right)
$ and $\beta_{s}=g\left(  n_{r}/\gamma_{sr}^{\ast}+n_{l}/\gamma_{cl}\right)  $
are two-photon interaction coefficients, $\gamma_{jk}=\Gamma-i\left(
\omega_{j}-\omega_{uk}\right)  $ $[j=p,c,s;$ $k=l,r]$ are complex one-photon
detunings, and $\gamma_{0}=\Gamma_{lr}+\left\vert \Omega_{c}\right\vert
^{2}/\gamma_{sr}^{\ast}+\left\vert \Omega_{c}\right\vert ^{2}/\gamma
_{pl}-i(\delta_{\omega}-\omega_{lr})$ is the complex two-photon detuning. Eqs.~(\ref{4wm-1}) are similar to those obtained by Harada \textit{et al.}~\cite{HaradaPRA013809} but here including the diffraction term $\pm i\nabla_{\perp}^{2}/(2q)$, required to explore the spatial
evolution of the fields.

We start with the simple case of a weak plane-wave probe $\Omega_{p}\left(
\mathbf{r}\right)  =f\left(  z\right)  e^{i\boldsymbol{k}_{\perp}^{p}%
\cdot\boldsymbol{r}_{\perp}}e^{i(k_{z}^{p}-k_{0}^{p})z}\ $directed at some
small angle $\theta\approx k_{\perp}^{p}/k_{0}^{p}\ll1$ relatively to the $z$
axis\ (Fig.~\ref{FIG1}). The generated Stokes field is then also a plane wave
$\Omega_{s}\left(  \mathbf{r}\right)  =g\left(  z\right)  e^{i\boldsymbol{k}%
_{\perp}^{s}\cdot\boldsymbol{r}_{\perp}}e^{i\left(  k_{z}^{s}-k_{0}%
^{s}\right)  z}$. Substituting into Eqs.~(\ref{4wm-1}), the phase-matching condition $\vec{k}_{\perp}^{s}=-\vec
{k}_{\perp}^{p}$ is readily obtained, and the resulting equations for $f$ and
$g $ are \cite{BoydPRA1981}
\begin{equation}\label{fg1}
f^{\prime}(z)  =Af(z)+Bg^{\ast}(z)e^{i2k_{\Delta}z}  \quad \textrm{and} \quad
g^{\prime\ast}(z)  =Cf(z)e^{-i2k_{\Delta}z}+Dg^{\ast}(z),
\end{equation}
where $2k_{\Delta}=\left(  2\boldsymbol{k}^{c}-\boldsymbol{k}^{p}%
-\boldsymbol{k}^{s}\right)  \cdot\hat{z}\thickapprox$ $k_{\perp}^{2}/q$ is the
phase mismatch scalar [see Figs.~\ref{FIG1}(b) and (c)].

Assuming $f\left(  0\right)  =1$ and $g\left(  0\right)=0$, we follow Ref.~\cite{BoydPRA1981} and find along the medium
\begin{subequations}
\label{a5}%
\begin{align}
f\left(  z\right)  e^{-ik_{\Delta}z}  & =\frac{A-\lambda_{2}}{\lambda
_{1}-\lambda_{2}}e^{\lambda_{1}z}-\frac{A-\lambda_{1}}{\lambda_{1}-\lambda
_{2}}e^{\lambda_{2}z},\\
g^{\ast}\left(  z\right)  e^{ik_{\Delta}z}  & =-\frac{C}{\lambda_{1}%
-\lambda_{2}}\left(  e^{\lambda_{1}z}-e^{\lambda_{2}z}\right)  ,
\end{align}
\end{subequations}
with the eigenvalues $\lambda_{1,2}=(A+D)/2 \pm [\left(  A-D-i2k_{\Delta
}\right)  ^{2}/4+BC]^{1/2}$. A similar analysis for the case of three-wave mixing was presented by Gavrielides \emph{et al.} \cite{GavrielidesJAP1987}.

In the limit where $\left\vert B\right\vert $ and $\left\vert C\right\vert $
are much smaller than $\left\vert A\right\vert $\ and $\left\vert D\right\vert
$, the solution is governed by independent EIT for the probe and Stokes fields
with little coupling between them. In the opposite limit, the fields
experience strong coupling, and the real part of the eigenvalues
$\lambda_{1,2}$ can be made positive and result in gain.

\section{Comparison with experiments\label{s_experiments}}

To verify our model, we have calculated the probe transmission as a function
of the two-photon detuning and compared it to the data published in Refs.~\cite{HaradaPRA013809,LettArxiv2007}. The Doppler effect due to the motion of
the thermal atoms is taken into account by averaging the FWM coefficients,
$Q=A,B,C,D$ in Eq.~(\ref{a3}), over the Doppler profile \cite{HapperPR1976}. Assuming nearly
collinear beams, the mean coefficients are%
\begin{equation}
\overline{Q}=\frac{1}{\sqrt{2\pi}v_{th}}\int duQ\left(  \omega_{p}%
+qu,\omega_{c}+qu\right)  \exp\left(  \frac{-u^{2}}{2v_{th}}\right)
,\label{a9}%
\end{equation}
where $v_{th}=k_{B}T/m,$ $T$ \ the cell temperature, and $m$ the atomic mass.

\begin{figure}[tb]
\centering \includegraphics[width=13cm] 
{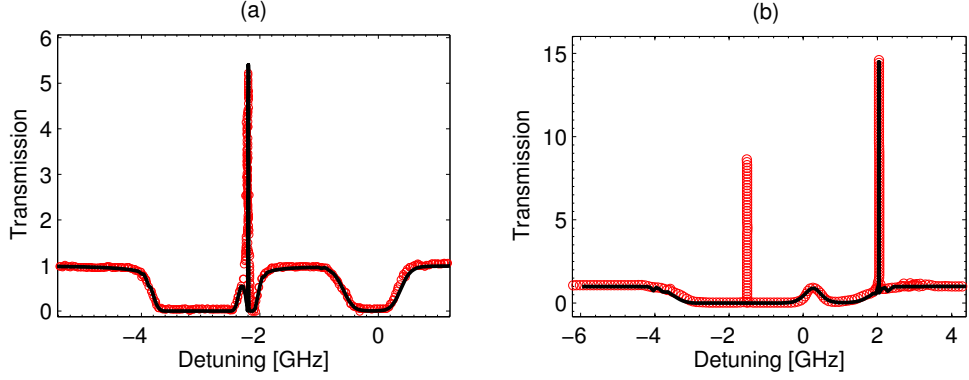}
\caption{Transmission spectra of FWM in (a) rubidium vapor and (b) sodium
vapor for a weak probe as a function of the two-photon detuning. The red
circles are experimental data from (a) Ref.~\cite{LettArxiv2007} and (b)
Ref.~\cite{HaradaPRA013809}. The black line is calculated from Eqs.~%
(\ref{a5})-(\ref{a9}) with the following parameters. For the rubidium
experiment: $\omega_{lr}=-3\ $GHz, $\Gamma_{rl}=5$ MHz, $\Omega_{c}=165$ MHz,
$\Gamma=5.7$ MHz, $N=1.9\times10^{12}$ cm$^{-3}$, $\Delta_{1p}=\omega
_{c}-\omega_{ur}=0.8$ GHz,$\ L=12.5$ mm, $T=150^{\circ}\ $C, and $k_{\perp
}/q=5.2\ $mrad. For the sodium experiment: $\omega_{lr}=1.777$ GHz,
$\Gamma_{rl}=1$ MHz, $\Omega_{c}=45$ MHz, $\Gamma$ MHz, $N=4.4\times
10^{11}$ cm$^{-3}$, $\Delta_{1p}=\omega_{c}-\omega_{ur}=2$ GHz, $L=45\ $mm,
$T=165^{\circ}\ $C, $k_{\perp}/q=4.5$ mrad.}%
\label{FIG3}%
\end{figure}

Figure \ref{FIG3} shows the transmission spectrum in (a) rubidium vapor and (b)
sodium vapor (cells length $L\simeq5$ cm). Here and in what follows, we characterize the resonance conditions by the one-photon detuning
$\Delta_{1p}=\omega_{c}-\omega_{ur}$ and the two-photon detuning
$\Delta_{2p}=\omega_{p}-\omega_{c}-\omega_{lr}$ (see Fig.~\ref{FIG1}a). Our model reproduces the
experimental spectra, including the Doppler-broadened absorption lines and the
gain peaks for both rubidium and sodium experiments. The missing peak in Fig.~\ref{FIG3}b is due to anti-Stokes generation not included in the model.%

\section{Diffraction manipulation by FWM\label{s_diffraction}}

We now concentrate on a specific choice of frequency detunings, for which the
phase dependency of the FWM process can be used to manipulate the diffraction
of the propagating probe and Stokes. To this end, we study the evolution of an
arbitrary image $F\left(  r\right)  $ imprinted on the probe beam, with the
boundary conditions $\Omega_{p}\left(  \boldsymbol{r}_{\perp},0\right)
=F\left(  r\right)  $ and $\Omega_{s}\left(  \boldsymbol{r}_{\perp},0\right)
=0.$ Our prime examples shall be the propagation of the image$\ $\emph{without
diffraction }or with\emph{\ reverse diffraction}, both of which while
experiencing gain.

\subsection{Image propagation}

We start by solving Eqs.~(\ref{4wm-1}) in the transverse Fourier basis,
\begin{subequations}
\label{Fk1}%
\begin{align}
\left(  \frac{\partial}{\partial z}+i\frac{k_{\perp}^{2}}{2q}-\bar{A}\right)
\Omega_{p}\left(  \boldsymbol{k}_{\perp},z\right)   & =\bar{B}\Omega_{s}%
^{\ast}\left(  \boldsymbol{k}_{\perp},z\right)  ,\\
\left(  \frac{\partial}{\partial z}-i\frac{k_{\perp}^{2}}{2q}-\bar{D}\right)
\Omega_{s}^{\ast}\left(  \boldsymbol{k}_{\perp},z\right)   & =\bar{C}%
\Omega_{p}\left(  \boldsymbol{k}_{\perp},z\right)  ,
\end{align}
\end{subequations}
where $\Omega_{p/s}\left(  \boldsymbol{k}_{\perp},z\right)  =\int dr_{\perp
}^{2}e^{-i\boldsymbol{k}_{\perp}\cdot\boldsymbol{r}_{\perp}}\Omega
_{p/s}\left(  \boldsymbol{r}_{\perp},z\right)  $, and the coefficients
$\bar{A}$, $\bar{B}$, $\bar{C}$, and $\bar{D}$ are Doppler averaged according
to Eq.~(\ref{a9}). We notice that Eqs.~(\ref{Fk1}) are identical to Eqs.~(\ref{fg1}) with $k_{\Delta}=0$ and with the substitutions $\bar{A}%
\rightarrow\bar{A}-ik_{\perp}^{2}/(2q)$ and $\bar{D}\rightarrow\bar
{D}+ik_{\perp}^{2}/(2q).$ The evolution of the probe and Stokes fields then
follows from Eq.~(\ref{a5}),
\begin{subequations}
\label{a12}%
\begin{align}
\frac{\Omega_{p}\left(  \boldsymbol{k}_{\perp},z\right)  }{\Omega_{p}\left(
\boldsymbol{k}_{\perp},0\right)  }  & =\frac{\bar{A}-ik_{\perp}^{2}%
/q-\lambda_{2}}{\lambda_{1}-\lambda_{2}}e^{\lambda_{1}z}-\frac{\bar{A}-ik_{\perp}^{2}/q-\lambda_{1}}{\lambda_{1}-\lambda_{2}%
}e^{\lambda_{2}z},\\
\frac{\Omega_{s}\left(  \boldsymbol{k}_{\perp},z\right)  }{\Omega_{p}\left(
\boldsymbol{k}_{\perp},0\right)  }  & =\frac{-\bar{C}}{\lambda_{1}-\lambda
_{2}}(e^{\lambda_{1}z}-e^{\lambda_{2}z}),
\end{align}
\end{subequations}
where
\begin{equation}
\lambda_{1,2}=\frac{\bar{A}+\bar{D}}{2}\pm\frac{1}{2}\sqrt{\left(  \bar
{A}-\bar{D}-i\frac{k_{\perp}^{2}}{q}\right)  ^{2}+4\bar{B}\bar{C}%
}.\label{a121}%
\end{equation}
We choose $|e^{\lambda_{2}z}|\gg|e^{\lambda_{1}z}|$ and obtain $\Omega
_{p}\left(  \boldsymbol{k}_{\perp},z\right)  =\Omega_{p}\left(  \boldsymbol{k}%
_{\perp},0\right)  e^{Z}$, where%
\begin{equation}
Z\equiv\lambda_{2}z+\log\left(  \frac{\bar{A}-ik_{\perp}^{2}/q-\lambda_{1}%
}{\lambda_{2}-\lambda_{1}}\right) \label{a13}%
\end{equation}
determines the changes in the spatial shape of the probe along its
propagation. $\operatorname{Re}Z$ is responsible for the $k_{\perp}%
$-dependency of the gain/absorption, and $\operatorname{Im}Z$ is responsible
for the $k_{\perp}$-dependency of the phase accumulation, that is, the
diffraction-like evolution.%

\begin{figure}[tb]
\centering \includegraphics[trim=0.362767in 0.000000in 0.262985in 0.000000in,width=8.3102cm]
{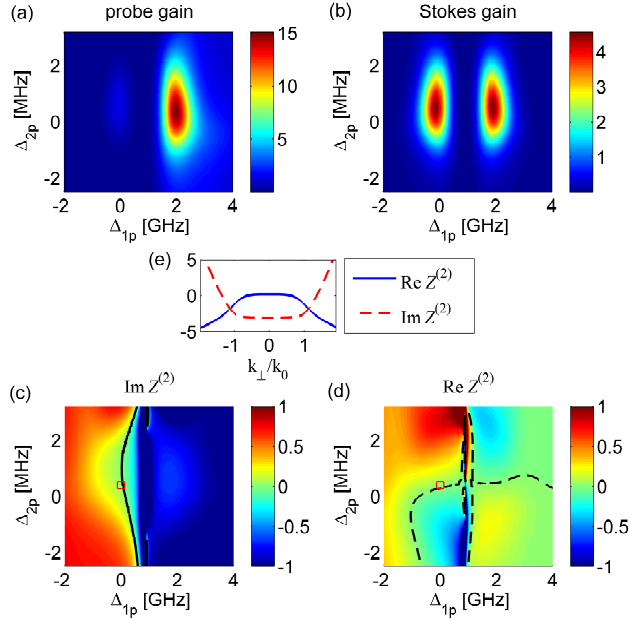}%
\caption{Numerical search for the detuning values that yield suppression\ of
paraxial diffraction and positive gain. This example uses the conditions of
the sodium system in Fig.~\ref{FIG3}. The colormaps as a function of the one-
and two- photon detunings are: (a) the Probe's gain, (b) Stokes' gain, (c)
$\operatorname{Im}Z^{(2)}$ of Eq.~(\ref{a14}), and\ (d) $\operatorname{Re}%
Z^{(2)}$. The contour $\operatorname{Im}Z^{(2)}=0$ is plotted in solid line in
(c). The contour$\ \operatorname{Re}Z^{(2)}=0$ is plotted in dashed line in
(d). (e) The exact propagation-exponent $Z$ [Eq.~(\ref{a13})]\ for the case
$\Delta_{2p}\thickapprox0.4$ MHz\ and $\Delta_{1p}\thickapprox0$\ [red square
in (c) and (d)]. At these detunings, $\operatorname{Re}Z^{(2)}%
=\operatorname{Im}Z^{(2)}=0$,\ and both the real and imaginary\ parts of
$Z$\ are constant for $k_{\perp}\ll k_{0}=$\ $40$ mm$^{-1}$. }%
\label{FIG4}
\end{figure} 

\subsection{Suppression of paraxial diffraction}

In general, the minimization of both the real and the imaginary $k_{\perp}$-dependencies of
$Z$ is required in order to minimize the distortion of the probe beam. To better understand the behavior of $Z$, we expand it in orders of
$k_{\perp}^{2}$. Taking the limit%
\begin{equation}
k_{\perp}^{2}\ll k_{0}^{2}=\min\left(  \frac{2qE^{2}}{\bar{A}-\bar{D}%
},2qE\right)  ,
\end{equation}
where $2E=[\left(  \bar{A}-\bar{D}\right)  ^{2}+4\bar{B}\bar{C}]^{1/2}$, we
write
\begin{equation}
{\small Z\thickapprox Z}^{(0)}+\frac{k_{\perp}^{2}}{2q}Z^{(2)}{\small +O}%
\left(  k_{\perp}^{4}\right)
\end{equation}
and find%
\begin{equation}
{\small Z}^{(0)}  {\small \thickapprox}\left(  \frac{\bar{A}+\bar{D}}%
{2}-E\right)  {\small z}, \quad \quad \quad
Z^{(2)}  =i\left(  \frac{\bar{A}-\bar{D}}{2E}z+\frac{\bar{A}-\bar{D}}%
{2E^{2}}-\frac{1}{E}\right)  .\label{a14}%
\end{equation}
The $k_{\perp}$-dependency, governed by $Z^{(2)}$, can be controlled through
the FWM coefficients $\bar{A}$, $\bar{B}$, $\bar{C}$, and $\bar{D}$ given in
Eq.~(\ref{a3}), by manipulating the frequencies of the probe and control fields
($\omega_{p},\omega_{c}$), the control amplitude $\Omega_{c}$, and the density
$N$.
\begin{figure}[tb]
\centering \includegraphics[width=7.783cm] {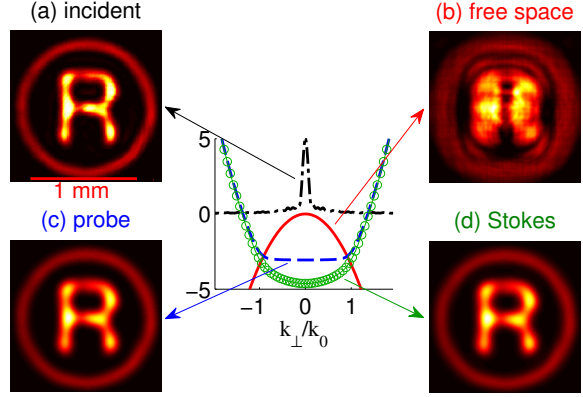}
\caption{Simulations demonstrating the suppression of paraxial diffraction by
FWM. (a) incident image, (b) after propagating in free space,\ (c) probe image
after propagating in the FWM medium, and (d) generated Stokes image. The
diffraction terms of the Probe (blue dashed line) and Stokes (green circles)
fields are compared with free-space diffraction (red line). The $k_{\perp}$
spectrum of the image is given for comparison (black dash-dot line). The probe
propagates in the cell with very little diffraction, and the Stokes'
distortion due to diffraction is reduced. The calculation is carried out in
the conditions highlighted in Fig.~\ref{FIG4} (detunings $\Delta_{2p}=0.4$ MHz
and $\Delta_{1p}=0$ and other parameters as in Fig.~\ref{FIG3}b). The image is
about 1-mm wide (features area 0.025 mm$^{2}$) and the propagation distance 45
mm (equivalent to $\lesssim2$ Rayleigh lengths).}%
\label{FIG6}
\end{figure}

We demonstrate this procedure in Fig.~\ref{FIG4}, using for example the experimental conditions of the sodium
experiment, detailed in Fig.~\ref{FIG3}. First, we observe the gain of the probe and the Stokes fields in
Fig.~\ref{FIG4}a and \ref{FIG4}b, as a function of the one- ($\Delta_{1p}$) and
two- ($\Delta_{2p}$) photon detunings. The gain is achieved around the
two-photon resonance ($\Delta_{2p}\ \thickapprox0$), either when the probe is
at the one-photon resonance ($\Delta_{1p}\thickapprox0$) or the Stokes
($\Delta_{1p}\thickapprox\omega_{lr},$ here $\thickapprox2\ $GHz); the latter
exhibits higher gain, since the probe sits outside its own absorption
line.\textbf{\ }The real and imaginary parts of $Z^{(2)}$ are plotted in Figs.~\ref{FIG4}c and \ref{FIG4}d. When $\operatorname{Re}Z^{(2)}=0$ (dashed line),
the gain/absorption is not $k_{\perp}$-dependent, whereas when
$\operatorname{Im}Z^{(2)}=0$ (solid line), the phase accumulation along the
cell is not $k_{\perp}$-dependent. When both happen, $Z^{(2)}=0,$ and a probe
with a spectrum confined within the resolution limit $k_{\perp}\ll k_{0}$
propagates without distortion. The exact propagation exponent $Z$ as a
function of $k_{\perp}$ for the point $Z^{(2)}=0$ ($\Delta_{2p}\thickapprox
0.4$ MHz, $\Delta_{1p}\thickapprox0$) are plotted in Fig.~\ref{FIG4}e. As expected, both real (blue solid line) and imaginary (red
dashed line) parts of $Z$ are constant for $k_{\perp}$ $\ll k_{0}$ (deviation
of 1\% within $k_{\perp}{\footnotesize <}k_{0}/2$ and 0.1\% within $k_{\perp}$
${\footnotesize <}$ $k_{0}/4$). In the specific example of Fig.~\ref{FIG4}, the probe's gain is $\sim$1.4, the Stokes' gain is $\sim$ 4, and
$k_{0}\thickapprox40\ $mm$^{-1}$.

To illustrate the achievable resolution, we shall employ a conservative
definition for a characteristic feature size in the image in area units
$a=(2\pi/k_{\perp})^{2}$ [For example: for a Gaussian beam, $a^{1/2}$ shall be
twice the waist radius, and, for the field pattern $E=1+\cos(k_{\perp}x)\cos(k_{\perp}y),$ the pixel area is $a^{2}.$ The Rayleigh length is
$qa^{2}/8$]. Fig.~\ref{FIG6} presents numerical calculations of Eqs.~(\ref{a12}) in the conditions found above for a probe beam in the shape of the
symbol (R) with features of $a\approx0.025$ mm$^{2}$ (corresponding to
$k_{\perp}=k_{0}=40$ mm$^{-1}$). The propagation distance is $L=45$ mm,
equivalent to $\lesssim2$ Rayleigh distances as evident by the substantial
free-space diffraction. Indeed when $Z^{(2)}=0$, the FWM medium dramatically
reduces the distortion of the image due to diffraction. Note that the image
spectrum (black dashed-dotted line) lies barely within the resolution limit
and that the Stokes distortion due to diffraction is also reduced. We emphasize that direct numerical solutions of Eqs.~(\ref{4wm-1}) give exactly the same results. For the hot sodium system, the
required control-field intensity is on the order of 100 mW for beams with a
waist radius of a few mm, which is practically a plane wave on the length
scale of the image.

\subsection{Negative paraxial diffraction}

Another interesting application of diffraction manipulation is imaging by
negative diffraction, similar to the one proposed in Ref.~\cite{FirstenbergNP2009}. Using the same tools as above, one can find the
conditions for the reversal of paraxial diffraction, namely when
$\operatorname{Re}Z^{(2)}$ vanishes and $\operatorname{Im}Z^{(2)}=1$ (free
space diffraction is equivalent to $Z^{(2)}=-i$).

\begin{figure}[tb]
\centering \includegraphics[width=12cm] 
{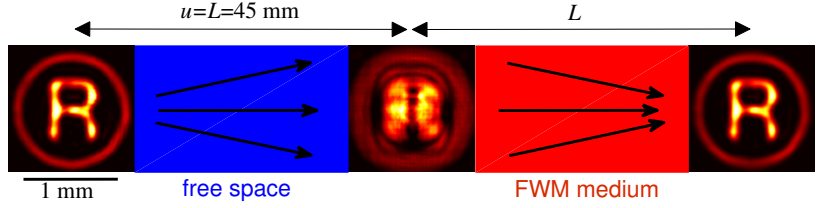}%
\caption{Demonstration of paraxial lensing with FWM. The negative-diffraction
medium of length $L$ focuses an image located at a distance $u<L$ to a
distance $v$\ behind the cell, where $u+v=L$. In this example, $u=L=45$ mm
and$\ v=0$. The probe gain is 1.5. The parameters (for which $Z^{(2)}=i$) are
$\Delta_{2p}=0.4$ MHz, $\Delta_{1p}=-1.7$ GHz, and $N=4\times10^{12}$
cm$^{-3}$; other parameters are as detailed in Fig.~\ref{FIG3}b.}%
\label{FIG7}
\end{figure}

At these conditions, as demonstrated in Fig.~\ref{FIG7}, the FWM medium of length $L$ focuses the radiation from a point
source at a distance $u<L$ to a distance $v$ behind the cell, where $u+v=L.$
The mechanism is simple: each $k_{\perp}$ component of the probe accumulates
outside the cell the phase $-ik_{\perp}^{2}\left(  u+v\right)
/(2q)=-ik_{\perp}^{2}L/(2q)$ and inside the cell the the phase $ik_{\perp}%
^{2}L/(2q)$, summing up to zero phase accumulation. The probe image thus
'revives', with some additional gain, at the exit face of the cell.

\begin{figure}[tb]
\centering \includegraphics[width=12cm] 
{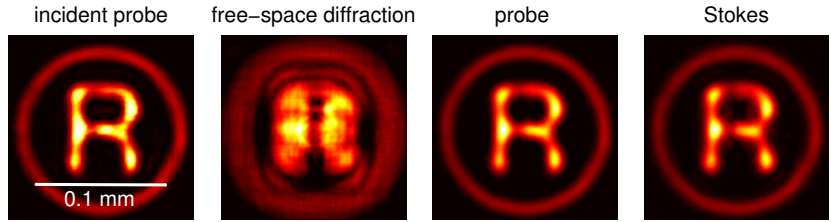}
\caption{Calculations for cold sodium atoms at a density of $N=10^{12}$ cm$^{-3}$ and negligible Doppler broadening. The image size is 0.1 mm
with 10-$\mu$m features. The propagation distance is 0.45 mm ($\sim$1.5
Rayleigh length).}
\label{HIGH_COLD}
\end{figure}

\section{Conclusions and discussion}

The suggested mechanism for manipulating the paraxial diffraction of light utilizes the $k_{\perp}$-dependency of the
four-wave mixing process and is thus fundamentally different than that suppressing the diffraction of spatial solitons in nonlinear media.
The inherent gain of the FWM process allows one to take advantage
of high optical-depths while avoiding absorption and, by that, achieving
higher resolution than with previous EIT-based schemes
\cite{FirstenbergPRL2009,FirstenbergNP2009}. As oppose to a recent proposal
incorporating FWM \cite{EversPRA2014}, our scheme does not require atomic
motion and is expected to work even more efficiently in its absence. We have
introduced a microscopic model for the FWM process, based on Liouville-Maxwell
equations and incorporating Doppler broadening, and verified it against recent
experimental results. The conditions for which the FWM process suppresses the paraxial diffraction were delineated. We have
demonstrated the flexibility of the scheme to surpass the regular
diffraction and reverse it, yielding an imaging effect while introducing gain.
Our proposal was designed with the experimental limitations in mind, and its
demonstration should be feasible in many existing setups.

The resolution limit $a^{-1}\propto k_{0}^{2}$ of our scheme (and thus the
number of 'pixels' $S/a$ for a given beam area $S$) is proportional to the
resonant optical depth. In practice, the latter can be increased either with
higher atomic density $N$ or narrower optical transitions. For example, using
a density of $N=5\cdot10^{12}$ cm$^{-3},$ 10 times higher than in the sodium
setup of Ref.~\cite{HaradaPRA013809}, the limiting feature area would be $250$ $\mu$m$^{2}$ ($k_{0}\thickapprox125$ mm$^{-1}$). As long as $NL=$\textit{const}$,$ the other parameters required for the suppression of diffraction remain the same.
At the same time, the reduced Doppler broadening in cold atoms media and in solids would substantially increase the resolution limit. Assuming
cold atoms with practically no Doppler broadening (and ground-state relaxation
rate $\Gamma_{lr}=100$ Hz), the same limiting feature of $250$ $\mu$m$^{2}$
can be obtained at a reasonable density of $10^{12}$ cm$^{-3}$. Finally, we
note that the best conditions for suppression of diffraction are not always
achieved by optimizing $Z^{(2)}$ alone (first order in $k_{\perp}^{2}%
/k_{0}^{2}$); In some cases, one could improve significantly by working with
higher orders. As demonstrated in Fig.~\ref{HIGH_COLD}, combining the aforementioned methods for resolution
enhancement with $N=10^{12}$ cm$^{-3}$ cold atoms, a resolution-limited
feature area as low as $100$ $\mu$m$^{2}$ with unity gain can be
achieved. Going beyond this resolution towards the $1-10$ $\mu$m$^{2}$-scale
for microscopy applications requires the lifting of the paraxial assumption in the analysis.

The FWM process conserves quantum coherence on the level of single photons, as was 
shown theoretically \cite{GlorieuxPRA2010} and experimentally
\cite{BoyerScience2008} by measuring spatial coherence (correlation) between
the outgoing probe and Stokes beams. An intriguing extension of our work would
thus be into the single-photon regime.
Specifically, the main limitation in the experiment of Ref.~\cite{BoyerScience2008} was the trade-off between focusing the beams to the
smallest spot possible while keeping the 'image' from diffracting throughout
the medium. Our scheme circumvents this trade-off by maintaining the fine
features of the image for much larger distances.

\section*{Acknowledgments}
We thank O. Peleg and J. Evers for helpful discussions.

\section*{Appendix: Steady-state solution}

Assuming control fields constant in time and space and much stronger than the
probe and Stokes fields, the steady-state solution of Eqs.~(\ref{e5})-(\ref{e6}) can be approximated to lowest orders in the weak fields
as $R_{\alpha,\alpha^{\prime}}\simeq R_{\alpha,\alpha^{\prime}}^{(0)}%
+R_{\alpha,\alpha^{\prime}}^{(1)},$where $R_{\alpha,\alpha^{\prime}}^{(0)}$ is
the zero-order and $R_{\alpha,\alpha^{\prime}}^{(1)} $ the first-order
steady-state solutions. We find $R_{\alpha,\alpha^{\prime}}^{(0)}$ from the
zero-order equations of motion%
\begin{align}
\frac{\partial}{\partial t}R_{l,l}^{(0)} &  =-2\operatorname{Im}(\Omega
_{c}^{\ast}R_{u,l}^{(0)})+\Gamma R_{u,u}^{(0)}+\Gamma_{l\leftrightarrow
r}(R_{r,r}^{(0)}-R_{l,l}^{(0)})\nonumber\\
\frac{\partial}{\partial t}R_{r,r}^{(0)} &  =-2\operatorname{Im}(\Omega
_{c}^{\ast}R_{u,r}^{(0)})+\Gamma R_{u,u}^{(0)}-\Gamma_{l\leftrightarrow
r}(R_{r,r}^{(0)}-R_{l,l}^{(0)})\nonumber\\
\frac{\partial}{\partial t}R_{u,u}^{(0)} &  =2\operatorname{Im}(\Omega
_{c}^{\ast}R_{u,l}^{(0)})+2\operatorname{Im}(\Omega_{c}^{\ast}R_{u,r}%
^{(0)})-2\Gamma R_{u,u}^{(0)}\nonumber\\
\frac{\partial}{\partial t}R_{r,l}^{(0)} &  =i\Omega_{c}^{\ast}R_{u,l}%
^{(0)}-i\Omega_{c}R_{u,r}^{\ast(0)}-i\omega_{lr}R_{r,l}^{(0)}-\Gamma
_{lr}R_{r,l}^{(0)}\nonumber\\
\frac{\partial}{\partial t}R_{u,l}^{(0)} &  =-i\Omega_{c}(R_{u,u}%
^{(0)}-R_{l,l}^{(0)})+i\Omega_{c}R_{r,l}^{(0)}+\gamma_{cl}R_{u,l}%
^{(0)}\nonumber\\
\frac{\partial}{\partial t}R_{u,r}^{(0)} &  =-i\Omega_{c}(R_{u,u}%
^{(0)}-R_{r,r}^{(0)})+i\Omega_{c}R_{r,l}^{\ast(0)}+\gamma_{cr}R_{u,r}^{(0)}%
\end{align}
by solving $(\partial/\partial t)R_{\alpha,\alpha^{\prime}}^{(0)}=0$. Under
the assumption $|\Omega_{c}/\omega_{lr}|\ll1,$ thus $R_{r,l}^{(0)}=0 $, we
obtain for the other elements%
\begin{align}
R_{l,l}^{(0)}  & =\frac{2A_{r}A_{l}+\Gamma A_{r}+\Gamma_{l\leftrightarrow
r}(A_{l}+A_{r}+\Gamma)}{X}\nonumber\\
R_{r,r}^{(0)}  & =\frac{2A_{r}A_{l}+\Gamma A_{l}+\Gamma_{l\leftrightarrow
r}(A_{l}+A_{r}+\Gamma)}{X}\nonumber\\
R_{u,u}^{(0)}  & =\frac{2A_{r}A_{l}+\Gamma_{1\leftrightarrow2}A_{r}%
+\Gamma_{l\leftrightarrow r}A_{l}}{X}\nonumber\\
R_{u,l}^{(0)}  & =i\frac{\Omega_{c}}{\gamma_{cl}}\frac{\Gamma\left(
\Gamma_{1\leftrightarrow2}+A_{r}\right)  }{X}\nonumber\\
R_{u,r}^{(0)}  & =i\frac{\Omega_{c}}{\gamma_{cr}}\frac{\Gamma\left(
\Gamma_{1\leftrightarrow2}+A_{l}\right)  }{X},
\end{align}
with the denominator%
\begin{align}
X  & =6A_{r}A_{l}+2\Gamma_{l\leftrightarrow r}\Gamma+A_{r}\left(  3\Gamma_{l\leftrightarrow r}+\Gamma\right)  +A_{l}\left(
3\Gamma_{l\leftrightarrow r}+\Gamma\right)  ,
\end{align}
and $A_{l/r}=\left\vert \Omega_{c}\right\vert ^{2}\operatorname{Im}%
[(\omega_{ul/ur}-\omega_{c}-i\Gamma)^{-1}]\ $the optical pumping rates.

To find $R_{\alpha,\alpha^{\prime}}^{(1)}$, we start from the first-order
equations of motion,%
\begin{align}
\frac{\partial}{\partial t}R_{l,l}^{(1)}  & =-2\operatorname{Im}\left[
\Omega_{p}^{\ast}e^{-i\delta_{k}z+i\delta_{\omega}t}R_{u,l}^{(0)}+\Omega
_{c}^{\ast}R_{u,l}^{(1)}\right]-\left(  \Gamma-\Gamma_{l\leftrightarrow r}\right)  R_{r,r}^{(1)}-\left(
\Gamma+\Gamma_{l\leftrightarrow r}\right)  R_{l,l}^{(1)}\nonumber\\
\frac{\partial}{\partial t}R_{r,r}^{(1)}  & =-2\operatorname{Im}\left[
\Omega_{s}^{\ast}e^{i\delta_{k}z-i\delta_{\omega}t}R_{u,r}^{(0)}+\Omega
_{c}^{\ast}R_{u,r}^{(1)}\right]-\left(  \Gamma+\Gamma_{l\leftrightarrow r}\right)  R_{r,r}^{(1)}-\left(
\Gamma-\Gamma_{l\leftrightarrow r}\right)  R_{l,l}^{(1)}\nonumber\\
\frac{\partial}{\partial t}R_{r,l}^{(1)}  & =i(\Omega_{s}^{\ast}R_{u,l}%
^{(0)}-\Omega_{p}R_{u,r}^{\ast(0)})e^{i\delta_{k}z-i\delta_{\omega}%
t} +i\Omega_{c}^{\ast}R_{u,l}^{(1)}-i\Omega_{c}R_{u,r}^{\ast(1)}-i\left(
\omega_{lr}-i\Gamma_{lr}\right)  R_{r,l}^{(1)}\nonumber\\
\frac{\partial}{\partial t}R_{u,l}^{(1)}  & =-i\Omega_{p}e^{i\delta
_{k}z-i\delta_{\omega}t}(R_{u,u}^{(0)}-R_{l,l}^{(0)}) +i\Omega_{c}(R_{r,r}^{(1)}+2R_{l,l}^{(1)}+R_{r,l}^{(1)})-\gamma_{cl}%
R_{u,l}^{(1)}\nonumber\\
\frac{\partial}{\partial t}R_{u,r}^{(1)}  & =-i\Omega_{s}e^{-i\delta
_{k}z+i\delta_{\omega}t}(R_{u,u}^{(0)}-R_{r,r}^{(0)}) +i\Omega_{c}(R_{l,l}^{(1)}+2R_{r,r}^{(1)}+R_{r,l}^{\ast(1)})-\gamma
_{cr}R_{u,r}^{(1)}.\label{cd-1}%
\end{align}
Eqs.~(\ref{cd-1}) are explicitly time-dependent, and we cannot directly solve for
$(\partial/\partial t)R_{\alpha,\alpha^{\prime}}^{(1)}=0$. Instead, we
introduce the new variables $P_{\alpha,\alpha^{\prime}}^{(1)}$ and
$N_{\alpha,\alpha^{\prime}}^{(1)}$ and rewrite Eqs.~(\ref{cd-1}) using
\begin{equation}
R_{\alpha,\alpha^{\prime}}^{(1)}=P_{\alpha,\alpha^{\prime}}^{(1)}e^{i \left[
\delta_{\omega}t-\delta_{k}z\right]  }+N_{\alpha,\alpha^{\prime}}%
^{(1)}e^{-i\left[  \delta_{\omega}t-\delta_{k}z\right]  },\label{RPN}%
\end{equation}
eliminating the explicit dependency on time. The steady-state solution is
obtained from the complete set of linear\ algebraic equations for the
variables $P_{u,l}^{(1)}$, $P_{u,r}^{(1)}$, $P_{r,l}^{(1)}$, $N_{u,l}%
^{\ast(1)}$, $N_{u,r}^{\ast(1)}$, $N_{r,l}^{\ast(1)}$, $P_{l,l}^{(1)}$, and
$P_{r,r}^{(1)}$,%
\begin{align}
0  & =\Omega_{c}^{\ast}P_{u,l}^{(1)}-\Omega_{c}N_{u,l}^{\ast(1)}+\Omega
_{p}^{\ast}R_{u,l}^{(0)}+i\left(  \Gamma-\Gamma_{l\leftrightarrow r}\right)  P_{r,r}^{(1)}+\left(
i\Gamma+i\Gamma_{l\leftrightarrow r}-\delta_{\omega}\right)  P_{l,l}%
^{(1)}\nonumber\\
0  & =\Omega_{c}^{\ast}P_{u,r}^{(1)}-\Omega_{c}N_{u,r}^{\ast(1)}-\Omega
_{s}R_{u,r}^{\ast(0)}+\left(  i\Gamma+i\Gamma_{l\leftrightarrow r}-\delta_{\omega}\right)
P_{r,r}^{(1)}+i\left(  \Gamma-\Gamma_{l\leftrightarrow r}\right)
P_{l,l}^{(1)}\nonumber\\
0  & =\Omega_{c}^{\ast}P_{u,l}^{(1)}-\Omega_{c}N_{u,r}^{\ast(1)}-\left(
\omega_{lr}-i\Gamma_{lr}+\delta_{\omega}\right)  P_{r,l}^{(1)}\nonumber\\
0  & =\Omega_{c}N_{u,l}^{\ast(1)}-\Omega_{c}^{\ast}P_{u,r}^{(1)}-\left(
\omega_{lr}+i\Gamma_{lr}-\delta_{\omega}\right)  N_{r,l}^{\ast(1)}-\Omega_{p}^{\ast}R_{u,r}^{(0)}+\Omega_{s}R_{u,l}^{\ast(0)}\nonumber\\
0  & =\Omega_{c}(P_{r,r}^{(1)}+2P_{l,l}^{(1)}+P_{r,l}^{(1)})+\left(
i\gamma_{cl}+\delta_{\omega}\right)  P_{u,l}^{(1)}\nonumber\\
0  & =\Omega_{c}^{\ast}(P_{r,r}^{(1)}+2P_{l,l}^{(1)}+N_{r,l}^{\ast
(1)})+\left(  i\gamma_{cl}^{\ast}+\delta_{\omega}\right)  N_{u,l}^{\ast
(1)} -\Omega_{p,l}^{\ast}(R_{u,u}^{(0)}-R_{l,l}^{(0)})\nonumber\\
0  & =\Omega_{c}^{\ast}(P_{l,l}^{(1)}+2P_{r,r}^{(1)}+P_{r,l}^{(1)}%
)-(i\gamma_{cr}^{\ast}-\delta_{\omega})N_{u,r}^{\ast(1)}\nonumber\\
0  & =\Omega_{c}(P_{l,l}^{(1)}+2P_{r,r}^{(1)}+N_{r,l}^{\ast(1)})-(i\gamma
_{cr}-\delta_{\omega})P_{u,r}^{(1)} +\Omega_{s,r}(R_{u,u}^{(0)}-R_{r,r}^{(0)})\label{ce-1}%
\end{align}

The exact solution of Eqs.~(\ref{ce-1}) is easily obtained but is unmanageable and bears no physical
intuition. Rather, we derive an approximate solution under the following assumptions:
\begin{enumerate}
\item The control and probe frequencies are near two-photon resonance,
$|\Delta_{2p}|=|\delta_{\omega}-\omega_{lr}|\ll\omega_{lr}$.
\item The ground-state population\emph{\ }relaxation is much slower than the
excited-to-ground relaxation, $\Gamma_{r\leftrightarrow l}\ll\Gamma$.
\item The optical pumping is much slower than the ground-state frequency
difference $\Omega_{c}^{2}/\Gamma\ll\omega_{lr}$.
\end{enumerate}
Under these assumptions, and taking the control Rabi frequency to be real
$\Omega_{c}=\Omega_{c}^{\ast}$, we solve Eqs.~(\ref{ce-1}) and obtain the coherences relevant to the evolution of the probe
and the Stokes [Eqs.~(\ref{e6})],
\begin{align}
iN_{u,l}^{(1)}  & =\left(  \frac{n_{l}}{\gamma_{pl}}+\frac{n_{r}}{\gamma_{cr}%
}-\frac{n_{l}/\gamma_{pl}+n_{r}/\gamma_{cr}^{\ast}}{\gamma_{pl}\gamma_{0}%
}\Omega_{c}^{2}\right)  \Omega_{p}+\frac{n_{r}/\gamma_{sr}^{\ast}+n_{l}/\gamma_{cl}}{\gamma_{pl}\gamma_{0}%
}\Omega_{c}^{2}\Omega_{s}^{\ast},\label{Nul1}\\
iP_{u,r}^{(1)}  & =\left(  \frac{n_{r}}{\gamma_{sr}}+\frac{n_{l}}{\gamma_{cl}%
}-\frac{n_{r}/\gamma_{sr}+n_{l}/\gamma_{cl}^{\ast}}{\gamma_{sr}\gamma
_{0}^{\ast}}\Omega_{c}^{2}\right)  \Omega_{p}+\frac{n_{l}/\gamma_{pl}^{\ast}+n_{r}/\gamma_{cr}}{\gamma_{sr}\gamma
_{0}^{\ast}}\Omega_{c}^{2}\Omega_{p}^{\ast}.\label{Pur1}%
\end{align}


\end{document}